%% file: ijcai24.tex
\documentclass{article}
\usepackage{ijcai24}

\usepackage{times}
\usepackage{soul}
\usepackage{url}
\usepackage[hidelinks]{hyperref}
\usepackage[utf8]{inputenc}
\usepackage[small]{caption}
\usepackage{graphicx}
\usepackage{amsmath}
\usepackage{amsthm}
\usepackage{booktabs}
\usepackage{algorithm}
\usepackage[switch]{lineno}
\usepackage{stfloats}

\usepackage{graphicx}
\usepackage{amsmath}
\usepackage{amssymb}
\usepackage{booktabs}
\usepackage{times}
\usepackage{epsfig}
\usepackage{algpseudocode}
\usepackage{multirow}
\usepackage{color}

\input{math_commands.tex}

\input{custom_commands.tex}

\title{A Training-Free Defense Framework for Robust Learned Image Compression}

\author{
    Myungseo Song ~~~~~~
    Jinyoung Choi ~~~~~~
    Bohyung Han \\
\affiliations
Computer Vision Laboratory, Seoul National University\\
\emails
\{micmic123, jin0.choi, bhhan\}@snu.ac.kr
}

\begin{document}

\maketitle



\input{sections/Abstract.tex}
\input{sections/Introduction.tex}
\input{sections/Related_works.tex}
\input{sections/Attack_on_compression.tex}
\input{sections/Defense.tex}
\input{sections/Experiments.tex}

\input{sections/Conclusion.tex}

\bibliographystyle{named}
\bibliography{egbib}

\input{sections/Appendix.tex}

\end{document}

%% file: math_commands.tex
\usepackage{amsmath,amsfonts,bm}

\def\eqref#1{equation~\ref{#1}}

\def\1{\bm{1}}

\DeclareMathAlphabet{\mathsfit}{\encodingdefault}{\sfdefault}{m}{sl}
\SetMathAlphabet{\mathsfit}{bold}{\encodingdefault}{\sfdefault}{bx}{n}



%% file: custom_commands.tex
\def\eg{\emph{e.g.}} 
\def\ie{\emph{i.e.}} 
\def\etal{\emph{et al.}}

\newcommand{\x}{\bm{x}}
\newcommand{\y}{\bm{y}}
\newcommand{\xhat}{\bm{\hat{x}}}
\newcommand{\yhat}{\bm{\hat{y}}}

\newcommand{\rom}[1]{(\MakeLowercase\expandafter{\romannumeral #1\relax})}

%% file: sections/Abstract.tex

\begin{abstract}
We study the robustness of learned image compression models against adversarial attacks and present a training-free defense technique based on simple image transform functions.
Recent learned image compression models are vulnerable to adversarial attacks that result in poor compression rate, low reconstruction quality, or weird artifacts.
To address the limitations, we propose a simple but effective two-way compression algorithm with random input transforms, which is conveniently applicable to existing image compression models. 
Unlike the na\"ive approaches, our approach preserves the original rate-distortion performance of the models on clean images.
Moreover, the proposed algorithm requires no additional training or modification of existing models, making it more practical.  
We demonstrate the effectiveness of the proposed techniques through extensive experiments under multiple compression models, evaluation metrics, and attack scenarios.
\end{abstract}

%% file: sections/Introduction.tex

\begin{figure}[ht]
  \centering
  \setlength\tabcolsep{1pt}
  \renewcommand{\arraystretch}{1}
  \scalebox{1}{
    \begin{tabular}{ccc}
      & \textsf{Clean} & \textsf{Perturbed} \\
      \rotatebox{90}{\hspace{13pt} \shortstack{\textsf{Input}}} &
      \includegraphics[width=.20\textwidth]{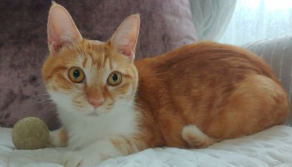} & 
      \includegraphics[width=.20\textwidth]{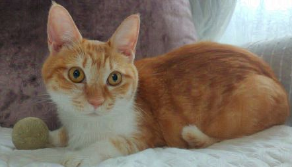} \\

      \rotatebox{90}{\hspace{-0.4pt} \shortstack{\textsf{w/o Defense}}} &
      \includegraphics[width=.20\textwidth]{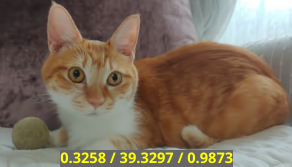} & 
      \includegraphics[width=.20\textwidth]{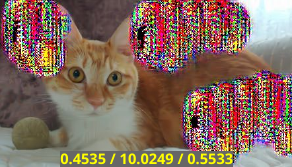} \\ 

      \rotatebox{90}{\hspace{1pt} \textsf{\shortstack{w/ Defense}}} &
      \includegraphics[width=.20\textwidth]{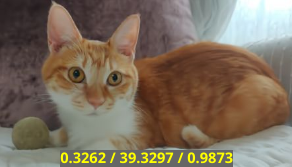} & 
      \includegraphics[width=.20\textwidth]{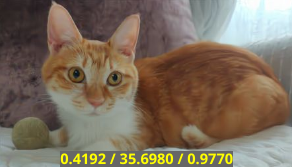} \\ 
    \end{tabular}
  }
  \caption{
      Demonstration of the vulnerability of learned image compression model to adversarial attacks and effectiveness of our defense method.
      The yellow annotations in each reconstructed image denote bits per pixel (bpp)/PSNR (dB)/MS-SSIM.
  }
\label{fig:demo}
\end{figure}

\begin{figure}[h]
  \centering
  \includegraphics[width=0.99\linewidth]{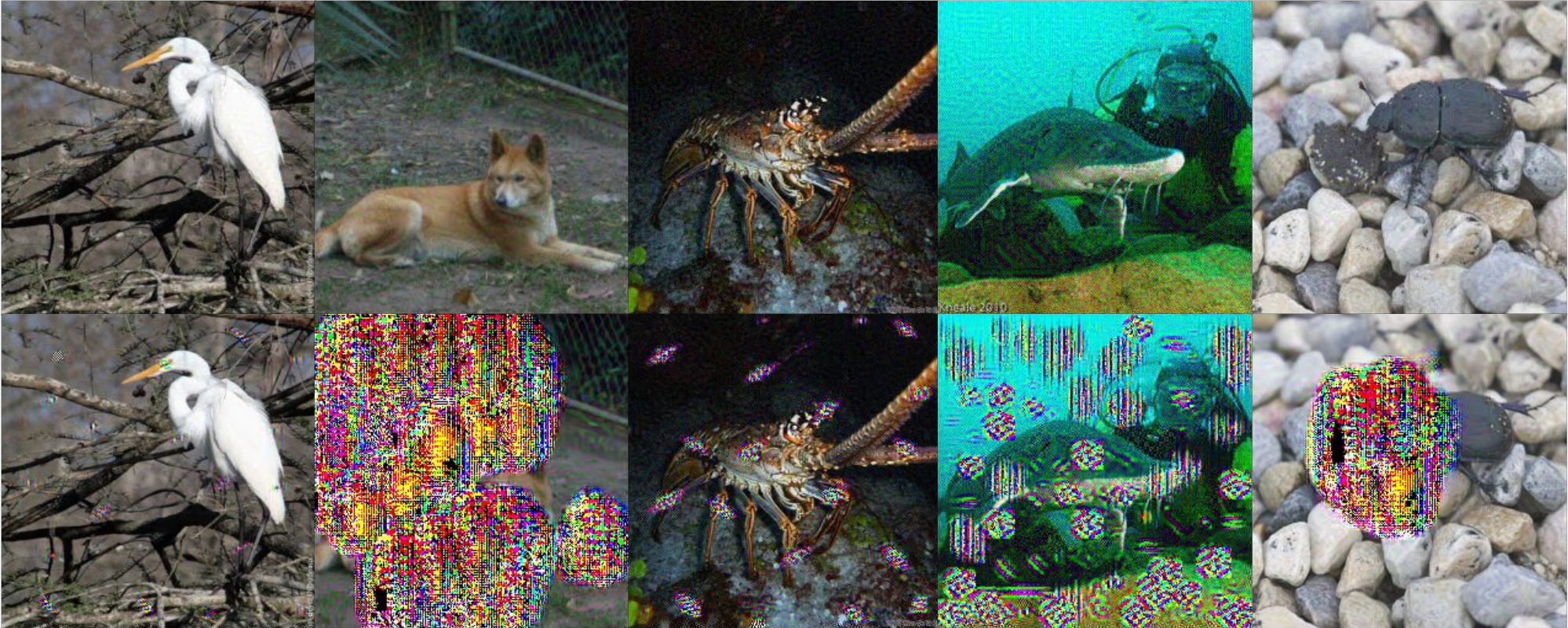}
  \caption{
      Examples of adversarially perturbed images (top) and corresponding reconstructed images (bottom).
  }
  \label{fig:qa_comparison}
\end{figure}

\begin{figure*}[t]
  \centering
  \setlength\tabcolsep{-1pt}  
  \begin{tabular}{cccc} 
    \includegraphics[width=.24\linewidth]{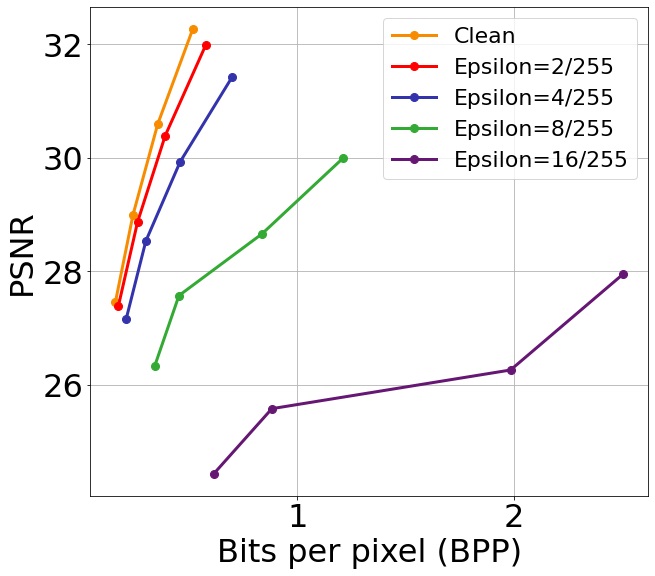} \ \ & 
    \includegraphics[width=.24\linewidth]{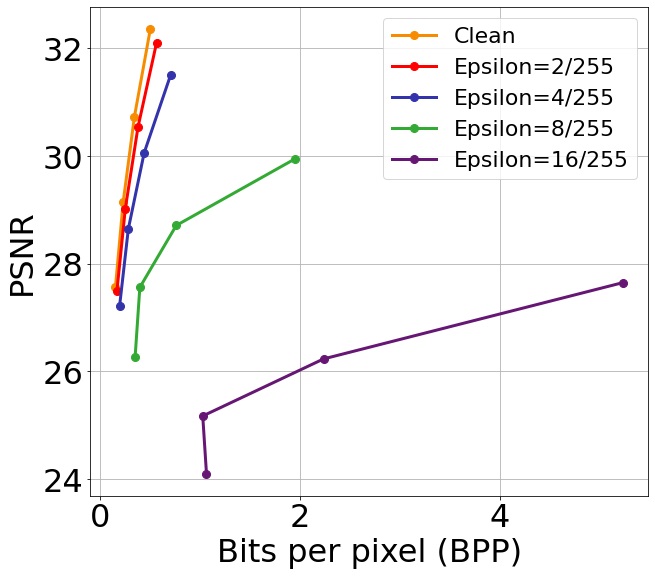} \ \ & 
    \includegraphics[width=.24\linewidth]{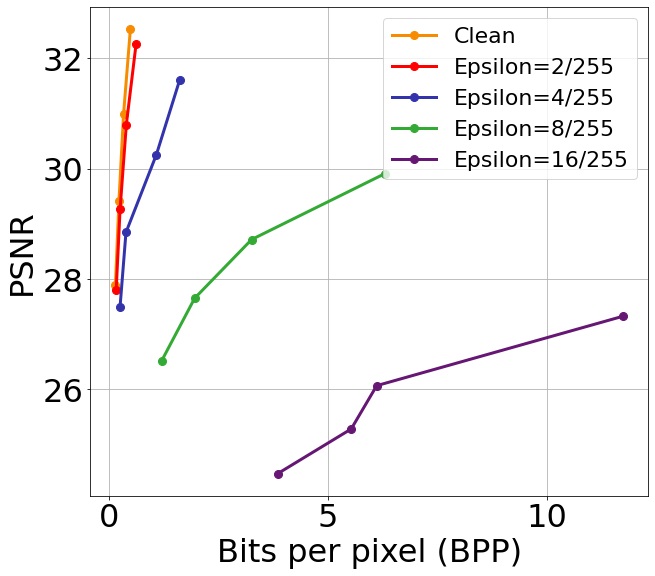} \ \ & 
    \includegraphics[width=.25\linewidth]{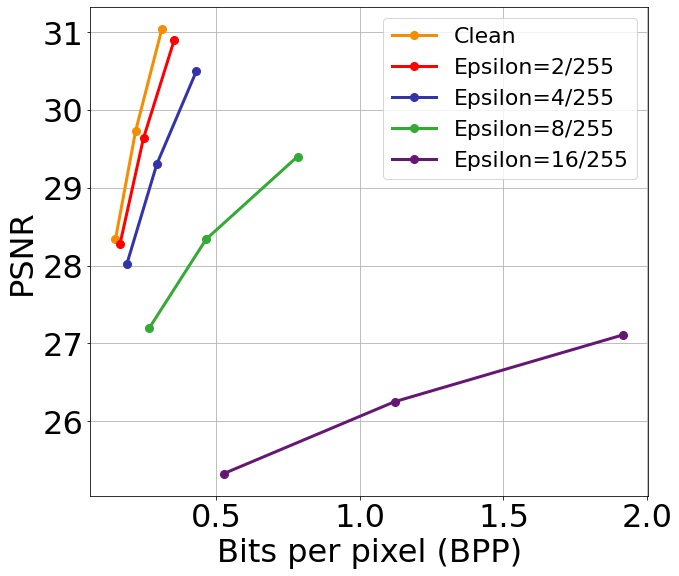} \\
    \includegraphics[width=.24\linewidth]{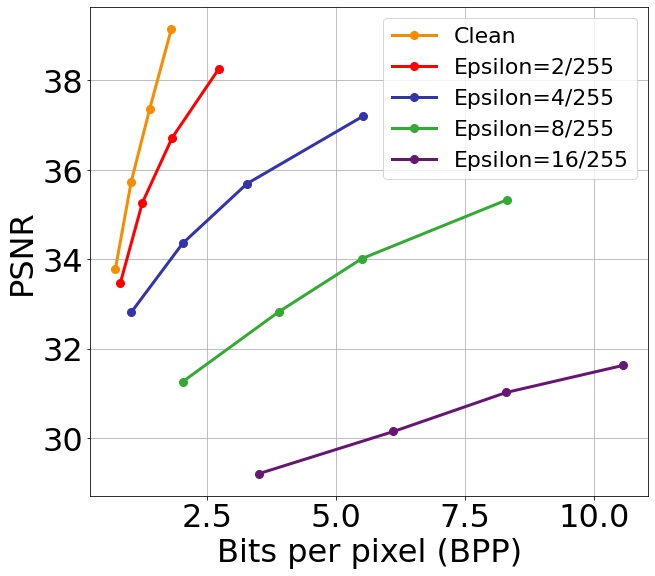} \ \ & 
    \includegraphics[width=.24\linewidth]{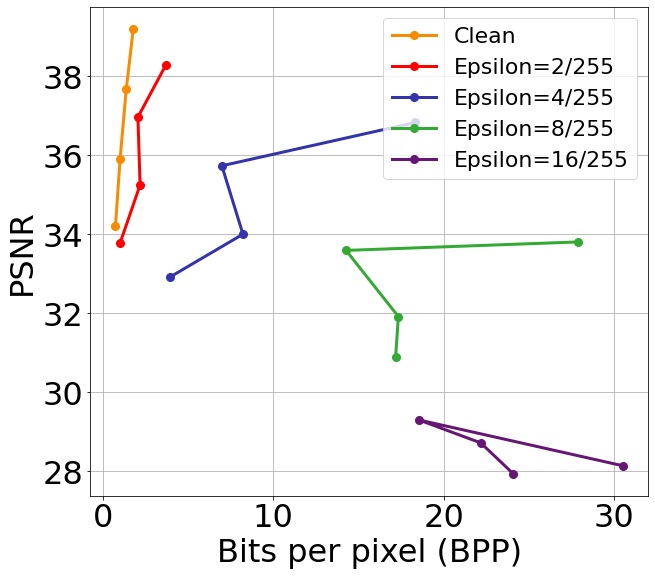} \ \ & 
    \includegraphics[width=.24\linewidth]{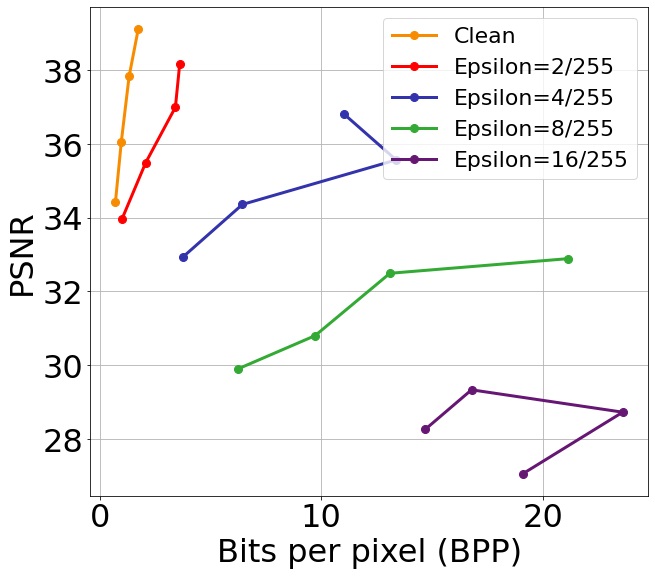} \ \ & 
    \includegraphics[width=.24\linewidth]{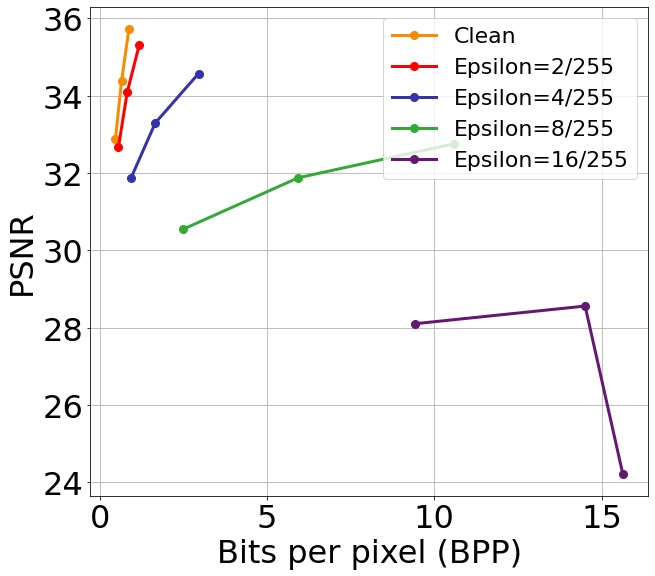} \\
    (a) SH & 
    (b) M\&S &
    (c) M\&S+C & 
    (d) Anchor \\
  \end{tabular}
  \caption{
    Results of adversarial attacks on image compression models for poor compression rates with various $\epsilon$ values for PGD algorithm.
    Top: results of low-bitrate models. 
    Bottom: results of high-bitrate models.
    Clean denotes the performance on clean (\ie, unperturbed) images.
  }
  \label{fig:vulnerability_epsilon}
\end{figure*}

\section{Introuduction}
\label{sec:introduction}
It is well-known that deep neural networks trained for image recognition are vulnerable to adversarial attacks~\cite{szegedy2014intriguing}.
By small and imperceptible perturbations on input images, the networks are easily deceived to behave for the intent of the attackers.
The performance of the models often drops significantly, which directly hampers the security and robustness of a whole system. 

As with other fields, adversarial attacks against learned image compression models are possible as well.
There are two feasible threats to lossy image compression, \ie, failure of bitrate reduction and severe distortion of decoded images. 
Figure~\ref{fig:demo} presents an example of perturbed image and corresponding decoded image by an image compression model with weird artifacts.
These limitations of image compression have far-reaching power affecting subsequent downstream tasks such as classification and detection.
In this respect, it is worth paying attention to the robustness of image compression models and their defense techniques against attacks. 

Compared to the recognition domains, the robustness of deep image compression models have not been studied comprehensively.
Some attack algorithms proposed for other tasks have turned out to be generalizable to image compression models~\cite{chen2023towards,liu2023manipulation,sui2023reconstruction,yu2023backdoor}.
However, defense techniques for image compression are not mature yet, and a na\"ive application of defense methods designed for other tasks may not work properly in image compression.

To enhance the robustness of image compression models, one can adopt approaches such as adversarial fine-tuning, a straightforward method suggested in~\cite{chen2023towards}.
However, this approach requires additional model training and consequently degrades the original compression performance of the models on normal, unattacked images.
Another defense strategy performs preprocessing on input images such as Gaussian blurring and bit depth reduction~\cite{xu2018feature}. 
However, these methods inevitably increase reconstruction errors of normal images due to the content loss caused by the preprocessing, as discussed in~\cite{yu2023backdoor}.

This work investigates the vulnerability of learned image compression models and introduces a training-free defense strategy.
We show that the performance of recent image compression models are easily harmed by basic attack algorithms in terms of rate and distortion.
To avoid these risks, we propose a simple yet effective image compression framework for defense.
Our framework improves the stability of compression performance to diverse adversarial attacks with negligible performance degradation on clean images.
It leverages input randomization in a safe way based on the self-supervised nature of the image compression problem.
Our approach is directly applicable to pretrained compression models without additional training, hence practical.
The effectiveness of our defense method against the attack is illustrated in Figure~\ref{fig:demo}.

The main contributions of this paper are summarized as \rom{1} the investigation of adversarial attacks on learned image compression models, \rom{2} the proposal of simple and effective defense techniques against the attacks, and \rom{3} the evaluation on the robustness of the proposed compression framework.

%% file: sections/Related_works.tex

\section{Related Works}
\label{sec:related_works}
This section briefly describes adversarial attack and defense methods in classificaiton and compression fields.

\subsection{Adversarial Robustness of Image Classification}
After Szegedy~\etal~\shortcite{szegedy2014intriguing} first showed the adversarial vulnerabilites of classifiers, several attack methods have been introduced, including FGSM~\cite{goodfellow2015explaining}, C\&W~\cite{carlini2017towards}, DeepFool~\cite{moosavi2016deepfool}, and PGD~\cite{madry2017towards}.
They share the key idea of adding minimal perturbations on an image iteratively towards the decision boundary of a classifier.
FDA~\cite{ganeshan2019fda} perturbs an image by disrupting the statistics of the intermediate features of a model.
For defense, the adversarial training, adding adversarial examples into training dataset, is a mainstream technique~\cite{goodfellow2015explaining,madry2017towards,tramer2018ensemble,kannan2018adversarial}.
As another line of research, \cite{guo2018countering,xie2018mitigating} attempts to reduce the chance of successful attacks by randomizing inputs while \cite{xu2018feature,samangouei2018defense} defend the models by denoising through optimization.

\subsection{Adversarial Robustness of Image Compression}
Learend image compression methods typically adopt autoencoder networks with auxiliary entropy models for probability distribution estimation of latent representations~\cite{balle2018variational,minnen2018joint,cheng2020learned}.
Adversarial attacks on image compression models are achieved by either increasing the bitstream lengths of latent representations or degrading the quality of decoded images.
Recently, researchers start to explore and investigate the adversarial robustness of image compression models.
For example, Chen and Ma~\shortcite{chen2023towards} corrupt the reconstruction quality of the models via distortion attack.
Although they leverage adversarial fine-tuning to address the vulnerabilites of the models, it leads to compression quality degradation of unattacked images.
Liu~\etal~\shortcite{liu2023manipulation} conduct transferring attacks~\cite{papernot2016transferability} using a JPEG-like substitution model in a black-box attack scenario.
Sui~\etal~\shortcite{sui2023reconstruction} propose a distortion attack algorithm with less perceptible perturbations, and Yu~\etal~\shortcite{yu2023backdoor} introduce a trigger injection model for backdoor attack.

%% file: sections/Attack_on_compression.tex

\begin{table}[t]
  \centering

  \scalebox{0.9}{
    \begin{tabular}{lrr}
      \toprule
      Model &
      Low bitrate & 
      High bitrate \\

      \midrule
      SH &
      5M & 
      12M \\ 

      M\&S &
      7M &
      18M \\

      M\&S+C &
      14M &
      26M \\

      Anchor &
      12M &
      27M \\

      \bottomrule
    \end{tabular}
  }
  \caption{
    The number of parameters of the compression models used in our experiments with respect to their target bitrates.
  }
  \label{tab:models_num_parameters}
\end{table}

\section{Adversarial Attack on Learned Image Compression}
\label{sec:attack}
This section presents the basic techniques of learned image compression and adversarial attacks on it.
Next, we discuss the vulnerability of image compression in diverse apsects.  

\subsection{Preliminaries}
\label{subsec:prelimianries}
The goal of lossy image compression is to minimize the bitstream length of an image while preserving the content in the image as much as possible.
Typically, a compression system consists of an encoder $E$, a decoder $D$, a quantizer $Q$, and an entropy model $P$.

Given a source image $\x$, $E$ transforms $\x$ to a latent representation $\y = E(\x)$, which is then converted to a quantized latent representation $\yhat = Q(\y)$.
To save $\yhat$, an entropy coding algorithm like the arithmetic coding~\cite{rissanen1981universal} encodes $\yhat$ into a bitstream with the probability distribution of $\yhat$ estimated by $P$.
The length of the resulting bitstream is approximately $-\log P(\yhat)$ with minor overhead hence is often used as a surrogate of the rate loss term.
For decoding, $D$ generates the reconstructed image $\xhat$ from the quantized latent representation $\yhat$, \ie, $\xhat = D(\yhat)$.
Given a distrotion metric $d(\cdot, \cdot)$ such as the mean squared error (MSE), the rate-distortion loss $\mathcal{L}_\text{RD}$ is given by the sum of the rate loss $\mathcal{L}_\text{rate} = -\log P(\yhat)$ and the distortion loss $\mathcal{L}_\text{dist} = d(\x, \xhat)$ as follows:
\begin{equation}
    \label{eqn:rd_loss}
    \mathcal{L}_\text{RD} = \mathcal{L}_\text{rate} + \lambda \mathcal{L}_\text{dist} = -\log P(\yhat) + \lambda d(\x, \xhat),
\end{equation}
where a Lagrangian multiplier $\lambda$ controls the rate-distortion trade-off.
Then, the objective of the image compression model is given by
\begin{equation}
  \label{eqn:rd_loss_objective}
  \min \mathbb{E}_{\x \sim  p_{\x}} \left [ \mathcal{L}_\text{RD} \right ].
\end{equation}

Our experiments use four pretrained lossy image compression models available at an open-source compression library~\cite{begaint2020compressai}: Scale Hyperprior (SH)~\cite{balle2018variational}, Mean \& Scale Hyperprior (M\&S)~\cite{minnen2018joint}, Mean \& Scale Hyperprior with context model (M\&S+C)~\cite{minnen2018joint}, and Anchor (Anchor)~\cite{cheng2020learned}. 
Table~\ref{tab:models_num_parameters} shows the number of parameters of the models.
Note that the models for high bitrates have more parameters than the low-bitrate counterparts.

\subsection{Attack Algorithm for Image Compression}
\label{subsec:pgd}

Among the adversarial attack strategies, we mainly adopt a famous optimization-based attack method, called the PGD algorithm~\cite{madry2017towards}.
To generate an adversarial example from a source image $\x$, PGD iteratively updates $\x$ with a step size $\alpha$ under the ${\ell}_\infty$-norm constraint of the maximum per-pixel perturbation $\epsilon$, which is given by
\begin{equation}
    \label{eqn:pgd_update}
    \x_{t+1} = \x_t + \alpha \cdot \text{sgn}(\bigtriangledown\mathcal{L}),
\end{equation}
where $\mathcal{L}$ denotes a task-specific loss and $\text{sgn}(\cdot) \in \{-1, 1\}$ is the sign function. 
Since compression models minimize the rate-distortion trade-off $\mathcal{L_\text{RD}}$, one can attack the model in terms of rate and distortion, for which the objective functions $\mathcal{L}$ are defined as $\mathcal{L}_\text{rate}$ and $\mathcal{L}_\text{dist}$, respectively.
It is also possible to employ the joint rate-distortion objective for attack by setting $\mathcal{L} = \mathcal{L}_\text{RD}$, but it makes the analysis more complex due to the conflicting properties of the two terms.
For the lossless image compression, only the rate loss is treated as a target since the source image content should be perfectly recovered.

\subsection{Results on Adversaries}
\label{subsec:results_on_adversaries}

\paragraph{Qualitative results} 
Figure~\ref{fig:qa_comparison} illustrates several adversaries of distortion attacks on M\&S and their corresponding reconstructed images.
The weird artifacts in the reconstructed images are easily induced by the attack, which shows the vulnerability of the model.

\paragraph{Quantitative results}
Figure~\ref{fig:vulnerability_epsilon} presents the results of adversarial attacks on four compression models with respect to the rate by varying the value of $\epsilon$ for the PGD algorithm. 
The larger $\epsilon$ is, the more performance degradation is observed consistently for all models.
Also, the high-bitrate models tend to be more vulnerable to the attacks than the low-bitrate ones.
This is partly because (i) the high-bitrate models with more parameters have more overfitting issues than the low-bitrate ones and (ii) the low-bitrate models have high reconstruction errors especially for high-frequency signals and hence tend to be robust to the adversarial noise given to input images.
The relationship between the model complexity and the vulnerability is discussed more in Appendix~\ref{sec:appendix_model_complexity}.
The result of distortion attack is presented in Appendix~\ref{sec:appendix_rate_attack}.
To mitigate these adversarial effects, appropriate defense techniques are requird.

%% file: sections/Defense.tex

\section{Defending Adversarial Attacks}
\label{sec:defense}
This section reviews the input randomization defense technique~\cite{xie2018mitigating} proposed for image classification, and discusses its limitations of direct application to image compression.
Then, we present our main idea of training-free defense technique for image compression models.

\subsection{Input Randomization for Image Classification}
The input randomization~\cite{xie2018mitigating} is a technique without training for mitigating the adversarial effects of image classification models.
It first defines a set of image transformations $\mathcal{T} = \{\tau_1, ..., \tau_n\}$, where $\tau_\theta$ is an image transformation (\eg, cropping).
For an input image $\x$, a transform $\tau_\theta$ is randomly sampled from $\mathcal{T}$ and the transformed image is given by 
\begin{equation}
\label{eq:transformation}
    \x^t = \tau_\theta(\x), ~~\text{where}~~ \tau_\theta \in \mathcal{T}.
\end{equation}
Then, $\x^t$ is fed to the classification model for prediction.
Specifically, \cite{xie2018mitigating} adopts resizing followed by zero padding for the transforms, $\mathcal{T}$.

The randomness provided by  random transforms improves the robustness of the model.
The attackers cannot perform precise inference due to the randomness; the attack is suboptimal because the attackers should consider all possible transforms if $n$ is sufficiently large.
Next, we describe how to apply it to image compression and its challenges.

\begin{figure}[t]
    \centering
    \begin{tabular}{ccc}
         \includegraphics[width=.28\linewidth]{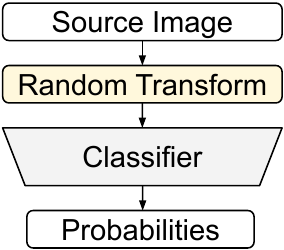} & 
         \includegraphics[width=.28\linewidth]{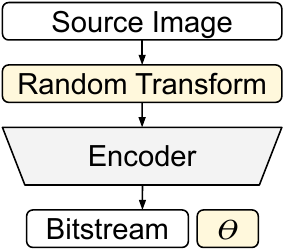} & 
         \includegraphics[width=.28\linewidth]{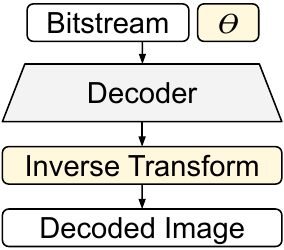} \\
        (a) & (b) & (c) \\
    \end{tabular}
    \caption{
       (a) Input randomization for image classification. 
       (b), (c) Input randomization for encoder and decoder of image compression.
    }
    \label{fig:input_randomization}
\end{figure}

\subsection{Input Randomization for Image Compression}
\label{subsec:input_random_compression}
To alleviate the adversarial effects on image compression models without additional training, we leverage the aforementioned input randomization technique~\cite{xie2018mitigating}.
Figure~\ref{fig:input_randomization} compares the input randomization in between image classification and image compression.

Suppose that we have a pretrained image compression model consisting of an encoder $E$, a quantizer Q and a decoder $D$.
To encode an input image $\x$, we first sample a transformation $\tau_\theta$ from $\mathcal{T}$ and transform $\x$ to get $\x^t$ as Equation~(\ref{eq:transformation}).
Then, we encode $\x^t$ instead of $\x$ as follows:
\begin{equation}
\label{eq:input_random_compression_encoding}
    \yhat = Q(E(\x^t)).
\end{equation}
The decoding is given by
\begin{equation}
\xhat^t = D(\yhat) ~~~~~~\text{and}~~~~~\xhat = \tau_\theta^{-1}(\xhat^t),
\label{eq:input_random_compression_decoding}
\end{equation}
where $\tau_\theta^{-1}$ is an inverse transform of $\tau_\theta$.
Note that $\mathcal{T}$ consists of (pseudo) invertible transforms for reconstruction and the additional cost to store the transform index $\theta$, $\log n$ bits, is negligible (about $4 \times 10^{-4}$ bpp in our experiments), compared to the bitstream of an image. 

Although such a na\"ive randomization approach improves adversarial robustness, the compression performance on normal images is degraded by some input transforms, which is further discussed below:
\begin{itemize}
    \item The cropping operations used in \cite{xie2018mitigating} are inappropriate due to incomplete reconstruction given by missing content.
    \item The transforms such as rotation, resizing and shifting have their corresponding inverse transforms, but the inversions are imperfect in general because of the information loss caused by the transforms, \ie, $\x \neq \tau^{-1}(\x^t)$.
    \item The zero padding operations utilized in \cite{xie2018mitigating} allow us to recover the original image, but the performance of the models would be degraded since the paddings lead to out-of-distribution images.
\end{itemize}

\begin{figure}[t]
    \centering
    \vspace{-5mm}
    \includegraphics[width=.7\linewidth]{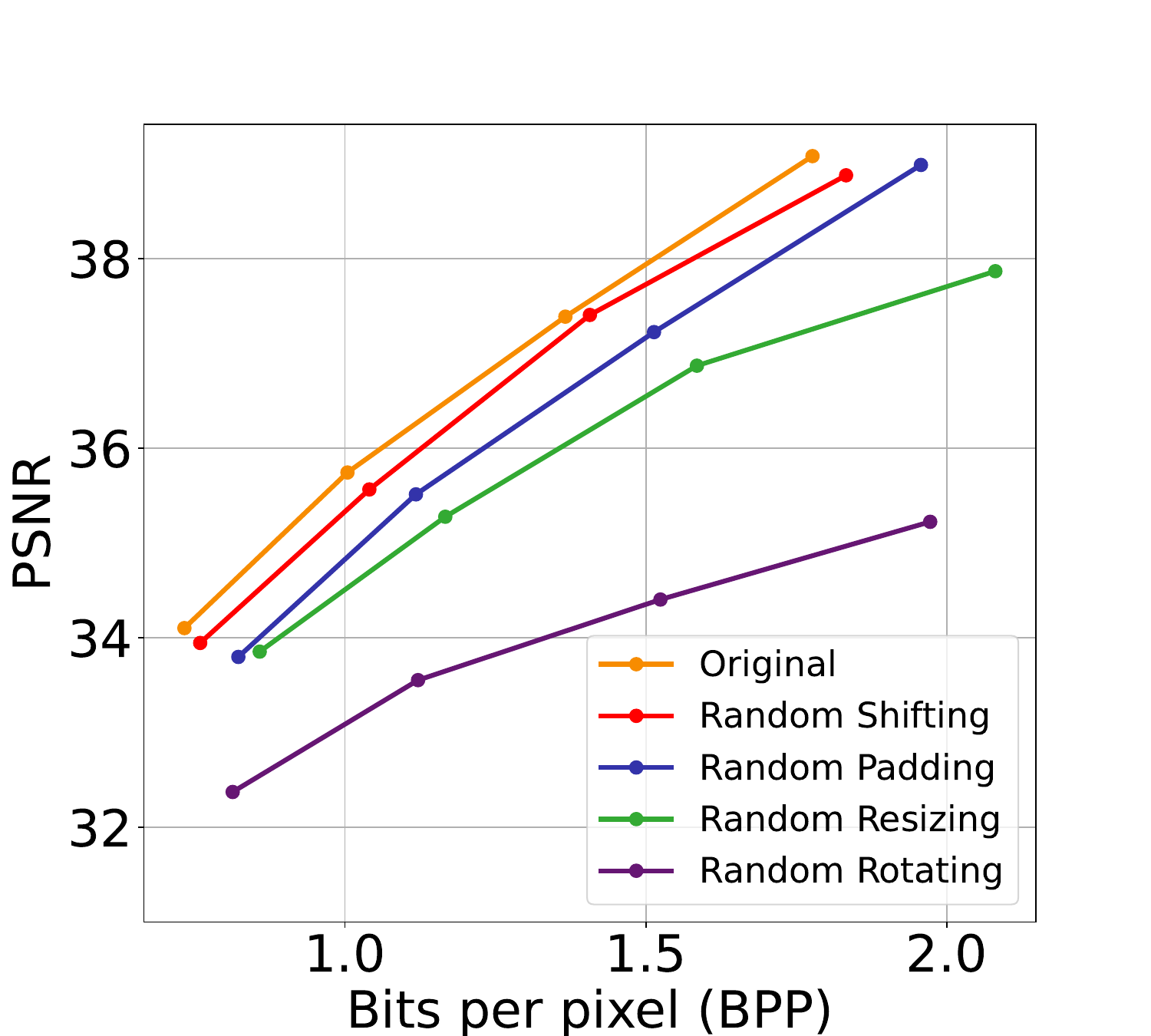}
    \caption{
       Performance degradation of an image compression model caused by a variety of input transforms.
    }
    \label{fig:naive_limitations}
\end{figure}

Figure~\ref{fig:naive_limitations} demonstrates the performance degradation of the image compression model~\cite{minnen2018joint} on clean images when various input transforms are applied.
Refer to Appendix~\ref{sec:appendix_input_transforms} for details.
It is not trivial to maintain the performance for these input transforms without additional training.

\subsection{Two-way Compression}
\label{subsec:two-way}
To defend against adversarial perturbations while preserving performance on clean images without additional model training, we propose a straightforward and training-free defense technique via two-way compression.
Our method is applicable to existing compression models without performance degradation on clean images by effectively leveraging the random transform.
In the framework, we select the better option out of two compression results of the original image and the randomly transformed image.
We summarize the encoding and decoding process of the proposed approach on Algorithm~\ref{alg:encoding} and Algorithm~\ref{alg:decoding}, respectively, where the entropy coding process is omitted for simplicity.

Our core idea is to choose the best compression strategy with the lowest loss value out of two different types of compression methods, which is feasible due to the availability of self-supervision in image compression.
The encoding process for an input image $\x$ is as follows.
First, we compute the rate-distortion loss of $\x$ given by encoding followed by decoding, without input transform.
The encoding and decoding are expressed as
\begin{equation}
    \yhat_1 = Q(E(\x)) ~~~~~~\text{and}~~~~~ \xhat_1 = D(\yhat_1),
\end{equation}
respectively.
Then, the rate-distortion loss of input image without transform is calculated by
\begin{equation}
    \mathcal{L}_1 = -\log_{2} P(\yhat_1) + \lambda d(\x, \xhat_1),
\end{equation}
where $d(\cdot, \cdot)$ is a distortion metric and $\lambda$ is a Lagrangian multiplier.
Next, we compute the rate-distortion loss of $\x$ with the input randomization as described in Section~\ref{subsec:input_random_compression}.
The encoding and decoding with the random input transformation are given by Equation~(\ref{eq:transformation}) to (\ref{eq:input_random_compression_decoding}), but we redefine  the latent representation and reconstructed image as $\yhat_2$ and $\xhat_2$, respectively.
The rate-distortion loss of input image with the random transform is given by
\begin{equation}
    \mathcal{L}_2 = -\log_{2} P(\yhat_2) + \lambda d(\x, \xhat_2).
\end{equation}
Finally, we determine the optimal compression result $\yhat^*$ and use it as the encoding result, which is given by
\begin{equation}
    \yhat^* = \begin{cases} 
        \yhat_1, & \text{if $\mathcal{L}_1 < \mathcal{L}_2$}. \\ 
        \yhat_2, & \text{otherwise}.
    \end{cases}
\end{equation}
For reconstruction, we save the transform index $\theta^*$ yielding the better result.
The decoding process is similar to Equation~(\ref{eq:input_random_compression_decoding}) with an input of $\yhat^*$.

\begin{algorithm}[t]
    \caption{Encoding phase of two-way compression}
    \label{alg:encoding}
    \textbf{Require}: Pretrained image compression model of encoder $E$, decoder $D$, quantizer $Q$, and entropy model $P$. \\
    \textbf{Require}: Distortion metric $d(\cdot, \cdot)$, Lagrangian multiplier $\lambda$, and Image transform set $\mathcal{T} = \{\tau_1, ..., \tau_n\}$. \\
    \textbf{Input}: Source image $\x$. \\
    \textbf{Output}: Compressed latent representation $\yhat^*$ and transform index $\theta^*$.
    \begin{algorithmic} 
        \State 1. Compute the loss for encoding without transform: \\
            \hskip2em Encode: $\yhat_1 \leftarrow Q(E(\x))$. \\
            \hskip2em Decode: $\xhat_1 \leftarrow D(\yhat_1)$. \\
            \hskip2em Compute loss: $\mathcal{L}_1 \leftarrow  -\log_{2} P(\yhat_1) + \lambda d(\x, \xhat_1)$. 
        \State 2. Compute the loss for encoding with random transform: \\
            \hskip2em Sample $\tau_\theta \in \mathcal{T}$. \\
            \hskip2em Apply transformation: $\x^t \leftarrow \tau_\theta(\x)$. \\
            \hskip2em Encode: $\yhat_2 \leftarrow Q(E(\x^t))$. \\
            \hskip2em Decode: $\xhat^t \leftarrow D(\yhat_2)$. \\
            \hskip2em Apply inverse transformation: $ \xhat_2 \leftarrow \tau_\theta^{-1}(\xhat^t)$. \\
            \hskip2em Compute loss: $\mathcal{L}_2 \leftarrow  -\log_{2} P(\yhat_2) + \lambda d(\x, \xhat_2)$.
        \State 3. Select the latent representation with the lowest loss: \\
            \hskip2em \textbf{if } $\mathcal{L}_1 < \mathcal{L}_2$ \textbf{then} \\
            \hskip4em $\yhat^* \leftarrow \yhat_1$. \\
            \hskip4em $\theta^* \leftarrow 0$. \\
            \hskip2em \textbf{else} \\
            \hskip4em $\yhat^* \leftarrow \yhat_2$. \\
            \hskip4em $\theta^* \leftarrow \theta$. \\
            \hskip2em \textbf{end if}
    \end{algorithmic}
\end{algorithm}
\begin{algorithm}[t]
    \caption{Decoding phase of two-way compression}
    \label{alg:decoding}
    \textbf{Require}: Pretrained decoder $D$. \\
    \textbf{Require}: Image transform set $\mathcal{T} = \{\tau_1, ..., \tau_n\}$. \\
    \textbf{Input}: Compressed latent representation $\yhat^*$ and transform index $\theta^*$. \\
    \textbf{Output}: Reconstructed image $\xhat$.
    \begin{algorithmic}
        \State
            Decode: $\xhat^t \leftarrow D(\yhat^*)$. \\
            \textbf{if } $\theta^* = 0$ \textbf{then} \\
            \hskip2em $ \xhat \leftarrow \xhat^t$. \\
            \textbf{else} \\
            \hskip2em Apply the inverse transform: $ \xhat \leftarrow \tau_{\theta^*}^{-1}(\xhat^t)$. \\
            \textbf{end if}
    \end{algorithmic}
\end{algorithm}

The proposed two-way compression approach prevents the compression quality degradation on the original images while improving the adversarial robustness of the compression model. 
The original model performance ($\mathcal{L}_1$) is guaranteed at least because we select the better option for compression by the comparison between $\mathcal{L}_1$ and $\mathcal{L}_2$, 
This attribute is especially valuable for normal images.
Besides, the risk of the adversarial attack is mitigated by our input randomization scheme.
The proposed framework is simple, easy-to-implement, and even free from additional training.
Note that this strategy is feasible due to the nature of image compression problem, availability of self-supervision, \ie, the ground-truth that the model has to reconstruct is identical to the input image of the encoder.

\paragraph{Computational efficiency}
Our approach requires more computation in the encoding phase because it has to perform an extra encoding for the transformed image and decode two encoded images, for both the clean and transformed images.
However, learned compression algorithms involves several time-consuming modules other than encoders and decoders, such as entropy coders and entropy models.
Also, we can adopt a lightweight encoding algorithm in our encoding phase based on masked convolution instead of expensive serial prediction, which saves computational cost significantly, especially in high-performance models adopting autoregressive entropy models~\cite{minnen2018joint,cheng2020learned}.
This trick is frequently used for training models with heavy entropy models~\cite{minnen2018joint,minnen2020channel}.
Moreover, the costly operation of decoding the bitstream to $\yhat$ is not needed because $\yhat$ is already available.
The computational cost in the decoding phase is almost identical except the overhead of applying inverse transform, which is negligible in practice.
We present empirical results related to computational cost in Section~\ref{sec:experiements}.

\paragraph{Scalability}
We can generalize the proposed framework to $K$-way compression for more gain in robustness. 
We sample $K-1$ transforms from $\mathcal{T}$ and choose the best among the $K$ compression results including the one with no transform.
In this way, we easily scale-up the robustness of the model with trade-off between the robustness and encoding cost.
However, we show that $K=2$ (\ie, two-way compression) is practically sufficient in Section~\ref{sec:experiements}.

%% file: sections/Experiments.tex

\begin{figure*}[t]
  \centering
  \begin{tabular}{ccc}
  \includegraphics[width=.29\textwidth]{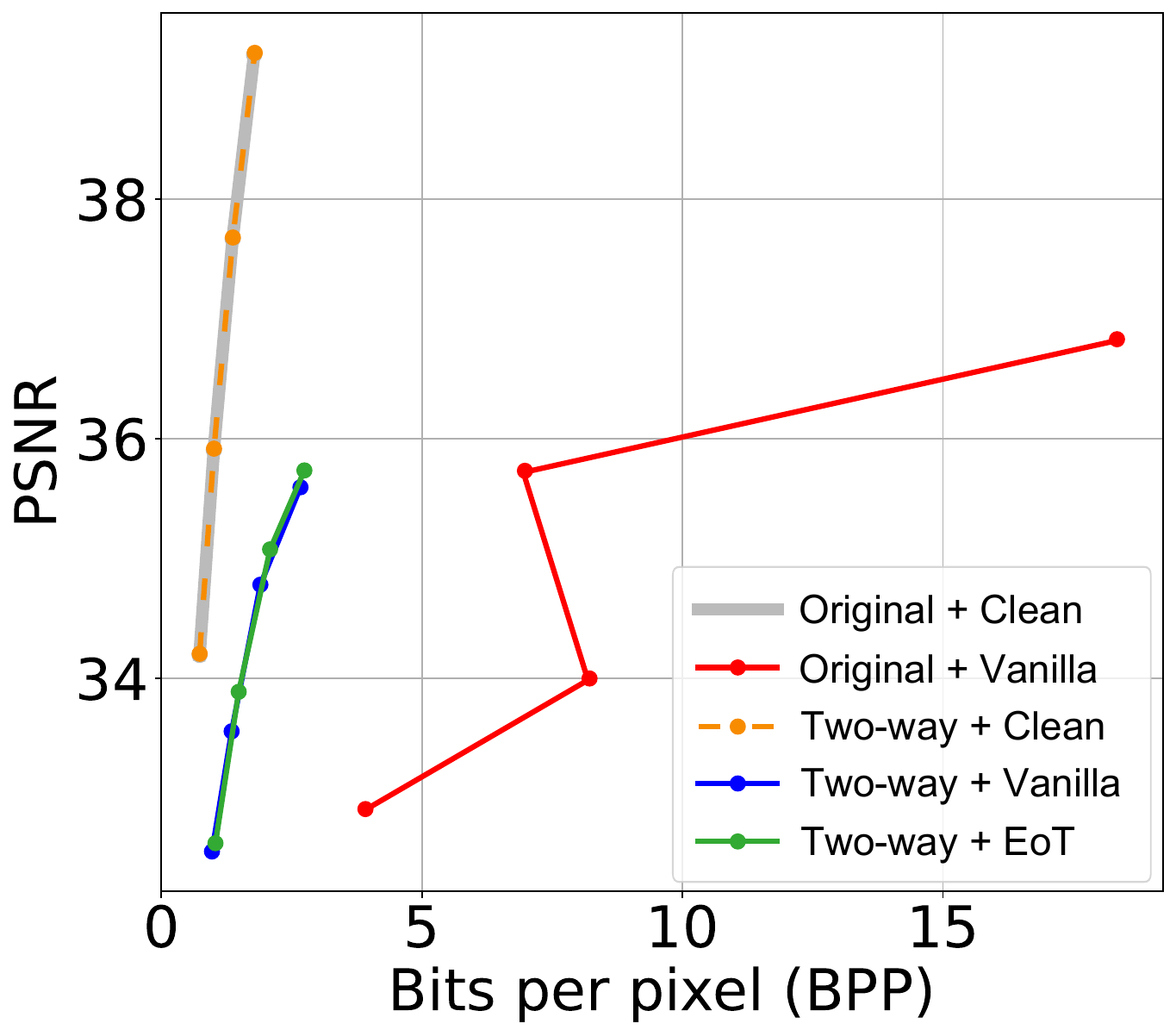} ~~ & 
  \includegraphics[width=.29\textwidth]{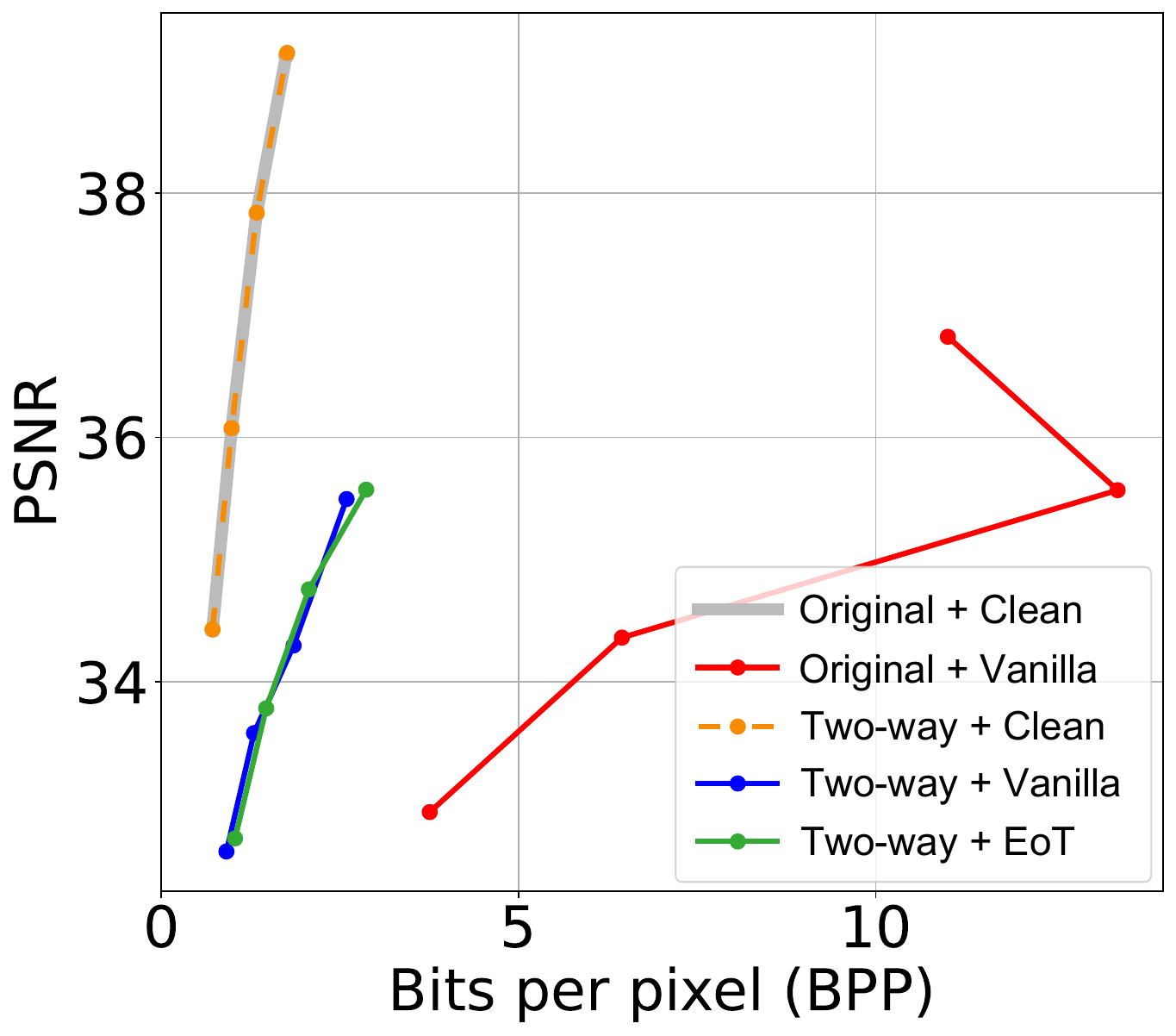} ~~ & 
  \includegraphics[width=.29\textwidth]{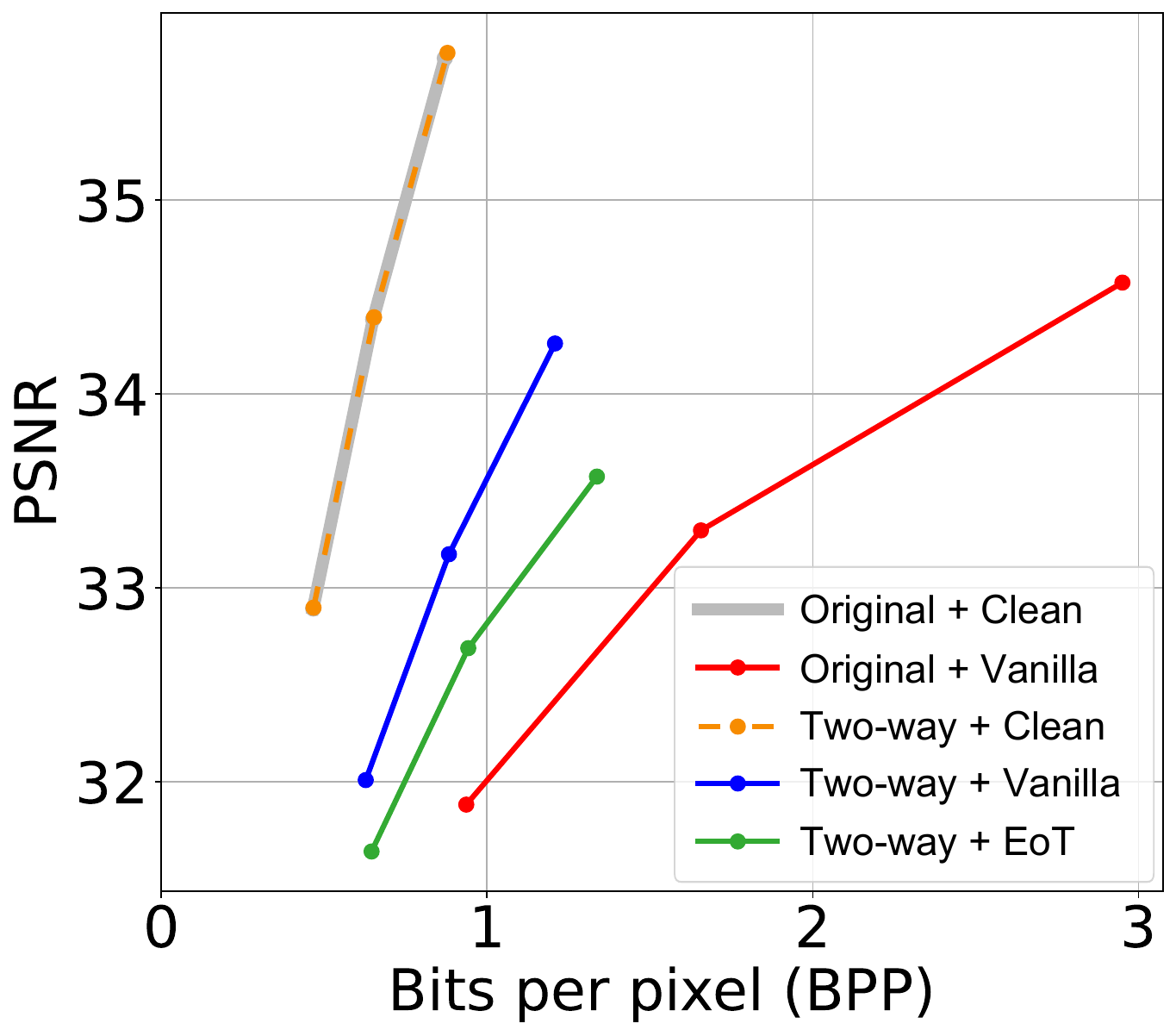} \\
  (a) M\&S ~~ & (b) M\&S+C ~~ & (c) Anchor
  \end{tabular}
  \caption{
    Rate-distortion performance of models without defense method (Original) and models with our defense method (Two-way) on clean images (Clean) and adversarial examples (Vanilla / EoT).
    Best viewed in color.
  }
  \label{fig:defense_main1}
\end{figure*}

\begin{figure}[t]
  \centering
  \includegraphics[width=0.9\linewidth]{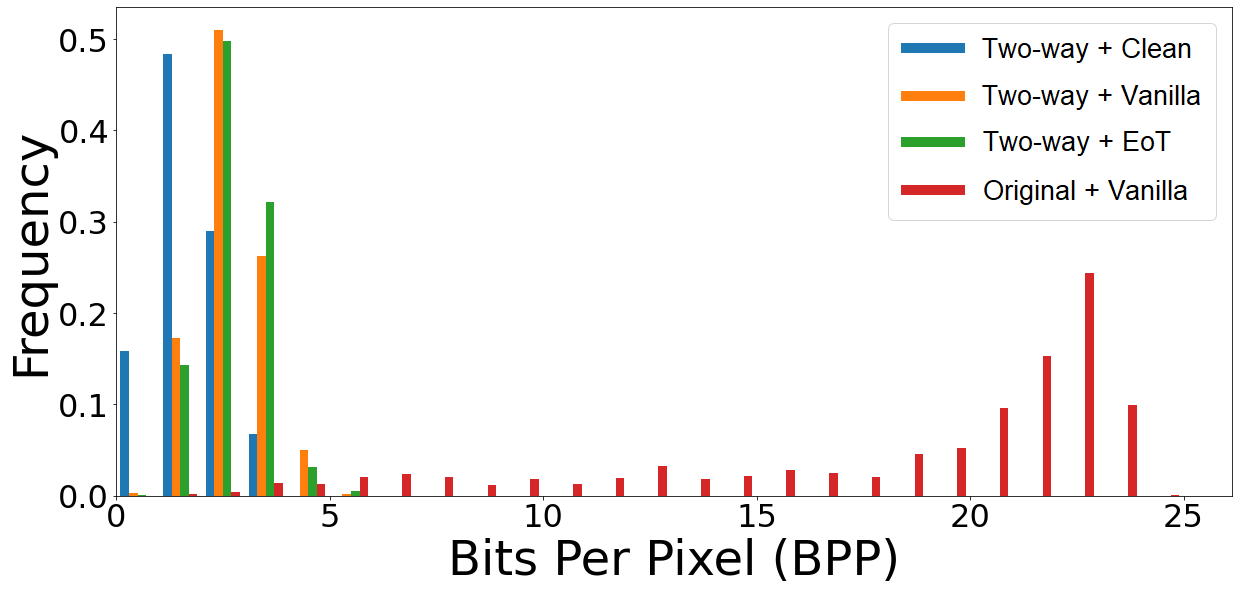}
  \caption{
    Bitrate histogram of test samples 
    under rate attacks.
  }
  \label{fig:defense_histogram}
\end{figure}

\begin{figure*}[t]
  \centering
  \begin{tabular}{ccc}
  \hspace{-0.3cm}
  \includegraphics[width=.29\textwidth]{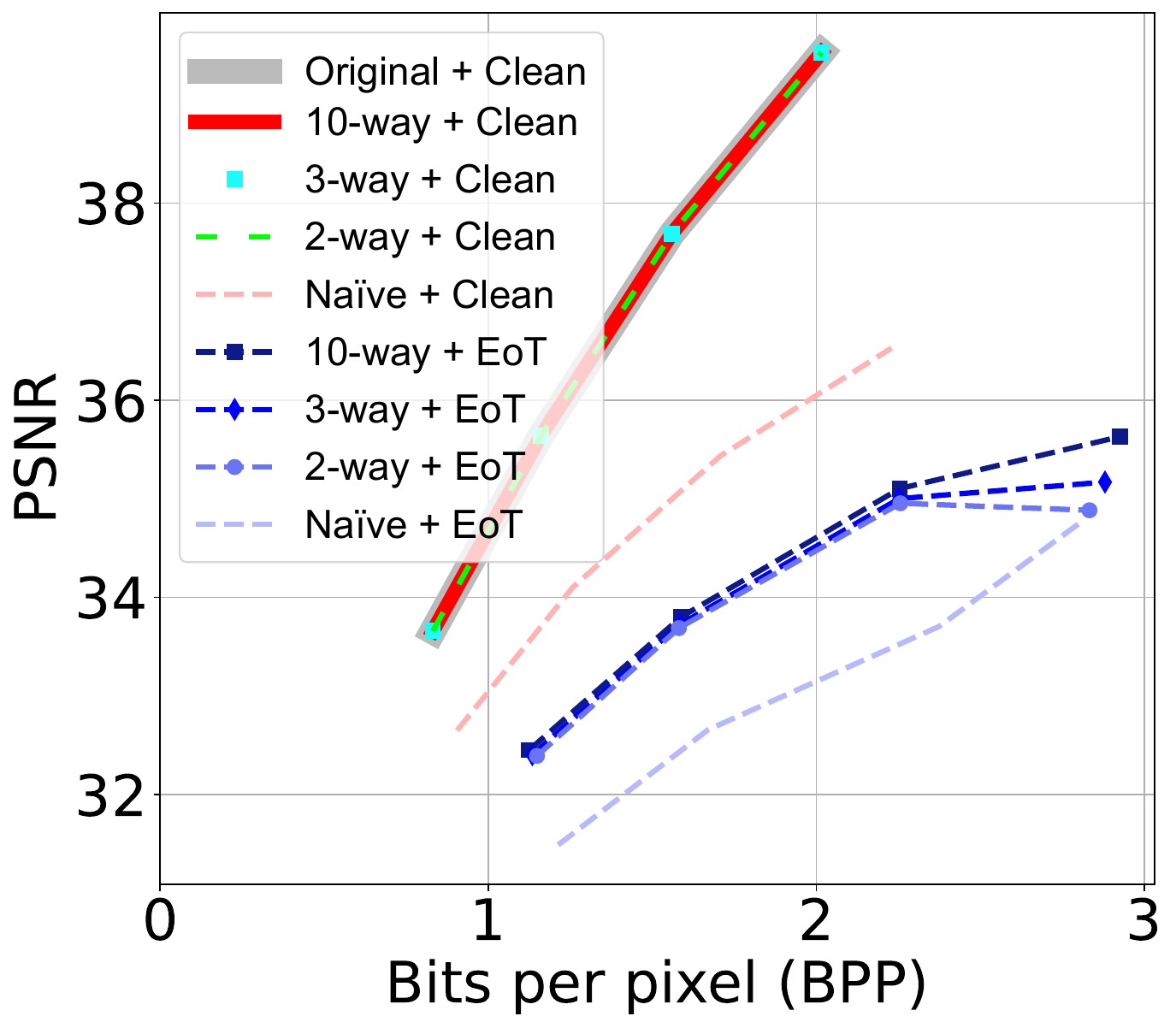} ~~ &
  \includegraphics[width=.29\textwidth]{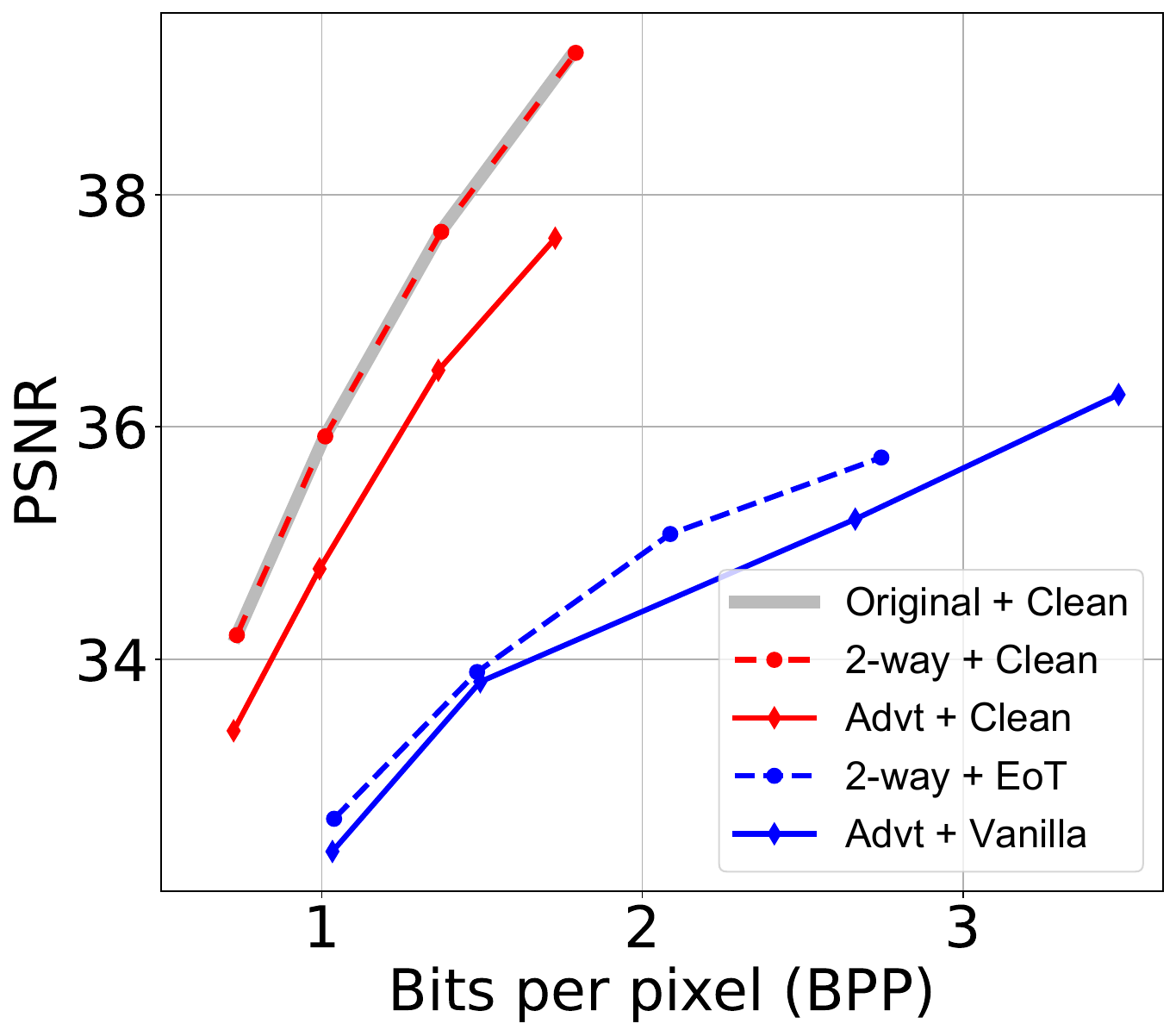} ~~ &
  \includegraphics[width=.29\textwidth]{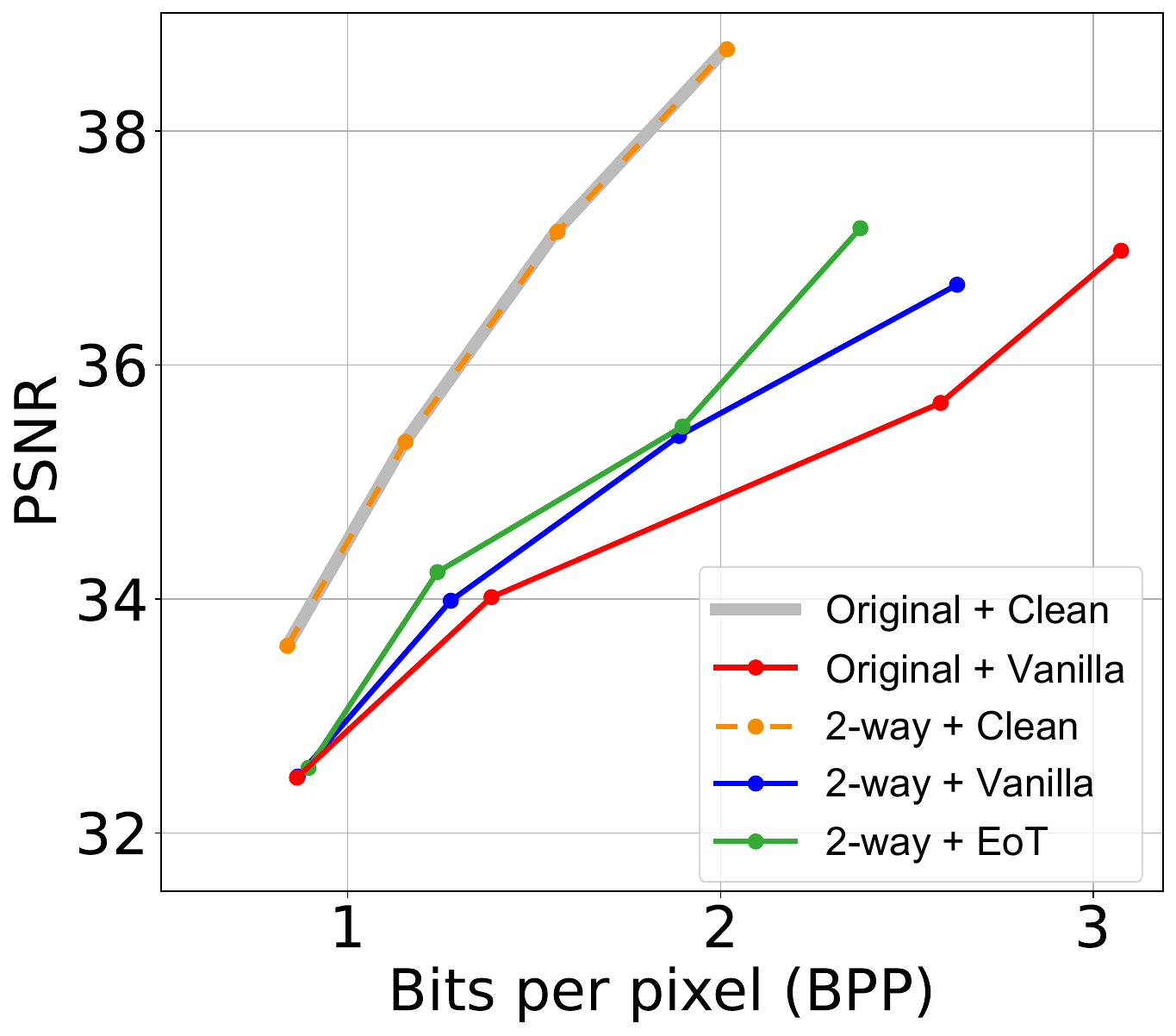} \\
  (a) ~~ & 
  (b) ~~ & 
  (c) 
  \end{tabular}
  \vspace{-2mm}
  \caption{
    Rate-distortion results of M\&S models for extensive studies. 
    (a) Results of $K$-way compression for multiple $K$ values and direct applicaiton of input randomization on image compression (Na\"ive).
    (b) Performance comparison between two-way compression and adversarial training (Advt).
    (c) Results of FDA attacks on original models and ones with our defense method.
  }
  \label{fig:defense_main2}
\end{figure*}

\section{Experiments}
\label{sec:experiements}
We now present the experimental results of the proposed defense framework.

\subsection{Experimental Setup}
The main experiments are conducted on 1000 validation images of $256\times256$ size randomly sampled from the ImageNet dataset~\cite{russakovsky2015imagenet}.
We use the pretrained high-bitrate models, Mean \& Scale Hyperprior (M\&S)~\cite{minnen2018joint}, Mean \& Scale Hyperprior with context model (M\&S+C)~\cite{minnen2018joint}, and Anchor (Anchor)~\cite{cheng2020learned}, as in Section~\ref{sec:attack}.
For image transform, we use the combinations of all elements in $\mathcal{T}$, which include (1) horizontal \& vertical flipping and rotating in multiples of 90 degrees ($8$ cases), (2) horizontal \& vertical stretching from 0 to 64 pixels ($65\times65 = 4225$ cases), and (3) horizontal \& vertical shifting from 0 to 64 pixels ($65\times65 = 4225$ cases).
These combinations result in $n = |\mathcal{T}| \approx 1.43 \times 10^8 $ transforms, where we only require less than 30 bits to store all possible indices. 

\paragraph{Attack scenarios}
We assume that the model weights are known to an attacker.
Our defense technique is tested in the following two scenarios depending on whether the attacker is aware of the existence of the defense method:
\begin{itemize}
  \item Vanilla attack: The attacker is not aware of the defense methods in the encoding algorithm, hence assumes input images are always fed to the model without modification (\ie, gray-box attack).
  \item Expectation over Transformation (EoT) attack: The attacker is aware of our two-way compression algorithm and transforms in $\mathcal{T}$, hence ideally aims to fool all the input transforms including the identity transform (\ie, white-box attack).
\end{itemize}
For the vanilla attack, we use the PGD algorithm as in Section~\ref{sec:attack} with $\alpha=2/255$, $\epsilon=4/255$, and 50 iterations.
The EoT attack~\cite{athalye2018synthesizing} is a strong white-box attack method for the two-way compression, which is often effective on the input randomization-based defense techniques~\cite{xie2018mitigating,guo2018countering} in classification.
Specifically, EoT attack randomly selects $24$ target transforms from $\mathcal{T}$ and average the losses of the target transforms at each optimization step of the PGD algorithm.

\subsection{Results}

\begin{table}[t]
    \centering
    \scalebox{0.9}{
      \begin{tabular}{lrr}
        \toprule
        Model & Original & Two-way \\ 
        
        \midrule
        M\&S    & 0.0219  & 0.0391 \\
        M\&S+C  & 0.7617  & 0.7952 \\
        Anchor & 0.7649  & 0.8437 \\ 
        
        \bottomrule
        \multicolumn{3}{c}{\vspace{-0.2cm}} \\
      \end{tabular}
    }
    \vspace{-3mm}
    \caption{
      Average encoding time of models in seconds.
    }
    \label{tab:defense_coding_time}
\end{table}

\paragraph{Main results}
Figure~\ref{fig:defense_main1} presents the performance of the proposed defense technique against rate attacks.
Overall, the proposed approach consistently improves the robustness of the models against the attacks.
In comparison to the severe performance degradation of original models by the attacks (`Original + Vanilla'), our method mitigates the adversarial effects (`Two-way + Vanilla' and `Two-way + EoT').
Furthermore, the performance of our method on clean images (`Two-way + Clean') is almost identical to the original one (`Original + Clean').
The attacks with multiple targets in EoT are more effective than the vanilla attack, which is highlighted in the Anchor model.  

Figure~\ref{fig:defense_histogram} visualizes the bitrate distribution of test samples for the highest bitrate models tested in the experiments for Figure~\ref{fig:defense_main1}(a).
Note that the results of our method exhibit low bpps by avoiding failure cases with high probability.
The histogram of rate-distortion loss is provided in Appendix~\ref{sec:appendix_histogram}.

\begin{figure*}[t]
    \centering
    \setlength\tabcolsep{1pt}
    \begin{tabular}{cccccc}
    \textsf{Clean}  & \textsf{Decoded} & ~~~ & \textsf{Perturbed}  & \textsf{Without defense}  & \textsf{With defense}  \\
      \includegraphics[width=.18\textwidth]{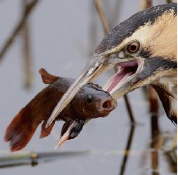} & 
      \includegraphics[width=.18\textwidth]{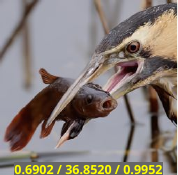} & ~~~ &
      \includegraphics[width=.18\textwidth]{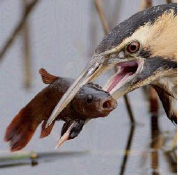} & 
      \includegraphics[width=.18\textwidth]{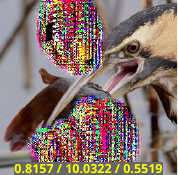} & 
      \includegraphics[width=.18\textwidth]{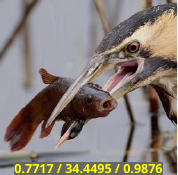} \\
      
      \includegraphics[width=.18\textwidth]{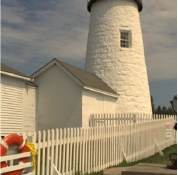} & 
      \includegraphics[width=.18\textwidth]{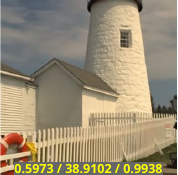} &~~~ &
      \includegraphics[width=.18\textwidth]{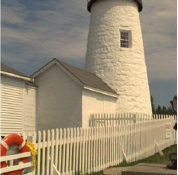} & 
      \includegraphics[width=.18\textwidth]{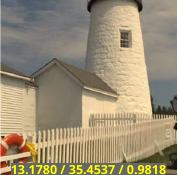} & 
      \includegraphics[width=.18\textwidth]{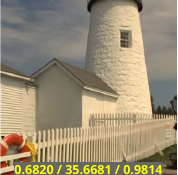} \\
    \end{tabular}
    \caption{
      Qualitative results of distortion attack (top) and rate attack (bottom).
      The first and second columns: original images and decoded results.
      The third and fourth columns: perturbed images and decoded results without our defense method.
      The last column: decoded results for the adversarial examples with our defense method.
      The yellow annotations denote bits per pixel (bpp)/PSNR (dB)/MS-SSIM.
    }
    \label{fig:defense_qualitative}
\end{figure*}

\paragraph{Scalability and na\"ive input randomization}
Figure~\ref{fig:defense_main2}(a) shows the defense results by varying $K$ in the $K$-way compression.
We used the Kodak dataset~\cite{kodak} and iteratively evaluated performance $40$ times for each sample.
Using a larger $K$ further improves the robustness of the model although the performance gains are saturated; two-way compression is sufficient for defense in practice.
Also, we test the na\"ive approach, applying the input randomization in image compression as described in Section~\ref{subsec:input_random_compression}, and report the results denoted by `Na\"ive' in Figure~\ref{fig:defense_main2}(a).
The difference between the na\"ive and two-way compression is that the former always encodes an input image with a random input transform while sharing $\mathcal{T}$.
Our defense framework clearly outperforms the na\"ive approach for both clean and perturbed images.

\paragraph{Comparison with adversarial training}
Figure~\ref{fig:defense_main2}(b) compares our defense method with adversarial training typically used in classification task.
We fine-tune the pretrained M\&S models using both the original images and the adversarial examples generated by FGSM with random initializations, following~\cite{wong2020fast}.
Our method outperforms the adversarial training in terms of the robustness to the attacks and the performance on clean images, even without training.

\paragraph{Generalizability}
To demonstrate the generalizability of the proposed defense method, we additionally test a feature-based attack method, feature disruptive attack (FDA)~\cite{ganeshan2019fda}.
For faster evaluation, we randomly sample 100 images from the test set and iteratively measure the performance 10 times for each sample.
As shown in Figure~\ref{fig:defense_main2}(c), our method consistently improves the robustness to FDA.

\paragraph{Encoding time} 
Table~\ref{tab:defense_coding_time} compares the encoding time of the original models and the models with our two-way compression technique on a single Titan Xp GPU.
The result shows the efficiency of our defense method.
Especially, the increase of encoding time is marginal for the high performance models (M\&S+C and Anchor) by utilizing masked convolutions for the loss computation as discussed in Section~\ref{subsec:two-way}.
Note that the extra cost for decoding is truly negligible and not tested.

\paragraph{Qualitative results} 
Figure~\ref{fig:defense_qualitative} qualitatively compares the impact of attacks and our defense methods along with the reconstructions of clean images.
Our defense methods decode the adversarial images as well as the clean ones, while maintaining a low bitrate that is competitive with the clean images.

%% file: sections/Conclusion.tex

\section{Conclusion}
\label{sec:conclusion}
We investigated the vulnerability of the learned image compression models and designed a simple yet effective defense method for image compression.
We observe that the performance of the recent image compression models can be easily harmed by the basic adversarial attacks in terms of rate and distortion.
The na\"ive defense approaches for image compression inevitably lead to performance degradation on clean images.
To address this, we present a robust defense framework for image compression that requires no additional training and preserves the original performance on clean images by exploiting the input randomization and characteristics of the self-supervised task.
The proposed algorithm computes the rate-distortion losses of the source image with random input transformation and identity transform, and chooses the best option in encoding. 
The combination of these two operations turns out to be effective while incurring a small amount of additional cost in the encoding phase.
Our framework is free from extensive training and modification of existing models, and can be easily integrated with various existing models.
This property is particularly desirable for robust image compression algorithms exposed to white-box adversarial attacks, where any trained models are vulnerable and unreliable.
We demonstrate the effectiveness of the proposed algorithm in white-box and gray-box attack scenarios and analyze the characteristics of our approach.

%% file: sections/Appendix.tex
\appendix
\section*{Appendix}

\section{Impact of Model Complexity to Robustness}
\label{sec:appendix_model_complexity}
\begin{figure}[t]
    \centering
    \includegraphics[width=.7\linewidth]{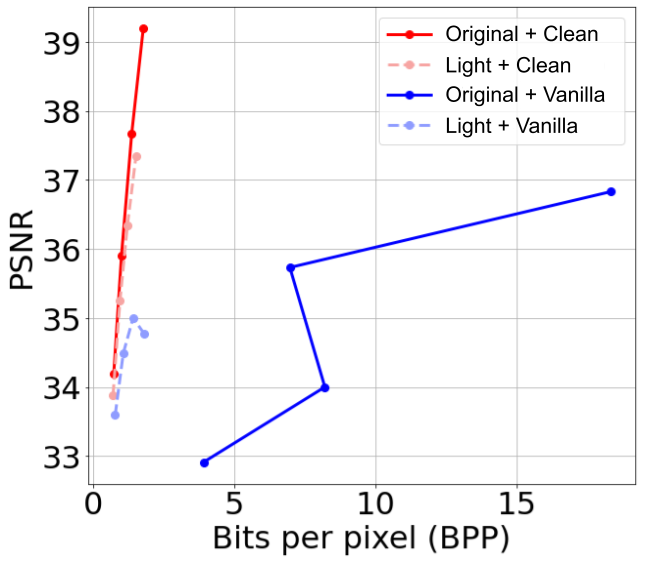}
    \vspace{-2mm}
    \caption{
       Rate-distortion results of original model and its lightweight verison with the half channel size.
    }
    \vspace{-2mm}
    \label{fig:size}
\end{figure}

To investigate the robustness of image compression models depending on the model complexity, we trained a lightweight variant of high-bitrate M\&S model, by halving its channel size.
Figure~\ref{fig:size} compares the results of the original model (18M parameters) and the lightweight model (7M parameters) under rate attacks.
While the model with higher capacity achieves slightly better performance on clean images, it suffers from significant failures on perturbed images. 
This implies that the model with higher capacity is more susceptible to adversarial attacks and rather overfitted.

\section{Results of Distortion Attacks}
\label{sec:appendix_rate_attack}
\begin{figure}[t]
    \centering
    \includegraphics[width=.7\linewidth]{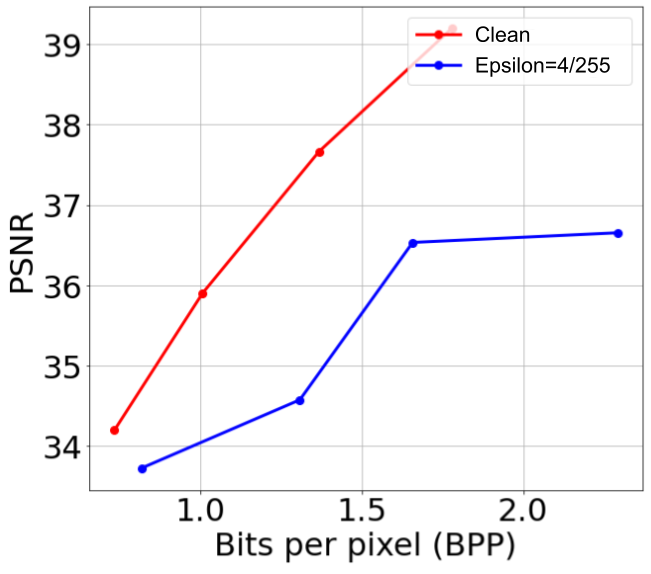}
    \vspace{-2mm}
    \caption{
        Rate-distortion result of distortion attacks.
    }
    \vspace{-2mm}
    \label{fig:vulnerability_etc}
\end{figure}

Figure~\ref{fig:vulnerability_etc} presents the result of distortion attack on M\&S model with $\epsilon=4/255$ for PGD algorithm.
The attacks for poor reconstruction quality successfully degraded the model performance.

\section{Details of Input Transforms}
\label{sec:appendix_input_transforms}
\begin{figure*}[t]
    \centering
    \includegraphics[width=1.0\linewidth]{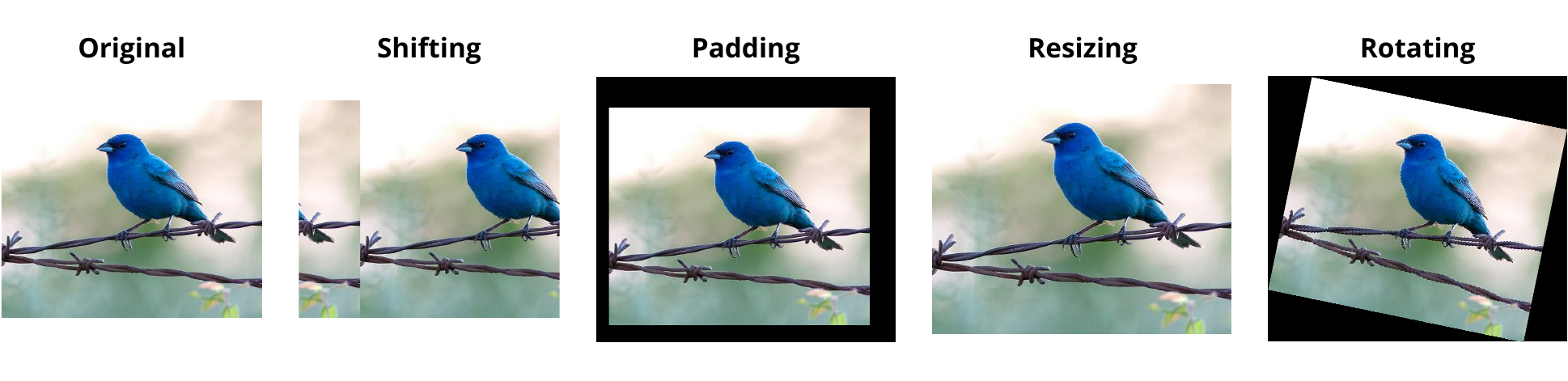}
    \vspace{-2mm}
    \caption{
        Examples of image transforms used in the experiments.
    }
    \label{fig:input_transforms}
\end{figure*}

This section explains the details of the image transforms used in the experiments for Figure~\ref{fig:naive_limitations} of the main paper.
The examples of the transformed images are illustrated in Figure~\ref{fig:input_transforms}.
For the image transforms, we use the operaitons including 
(1) horizontal and vertical shifting from 0 to 64 pixels,
(2) horizontal and vertical zero-padding from 0 to 32 pixels,
(3) horizontal \& vertical stretching from 0 to 64 pixels, and
(4) rotating from -10 to 10 degrees.

\section{Loss Histogram Under Attacks}
\label{sec:appendix_histogram}
\begin{figure*}[t]
    \centering
    \includegraphics[width=.7\linewidth]{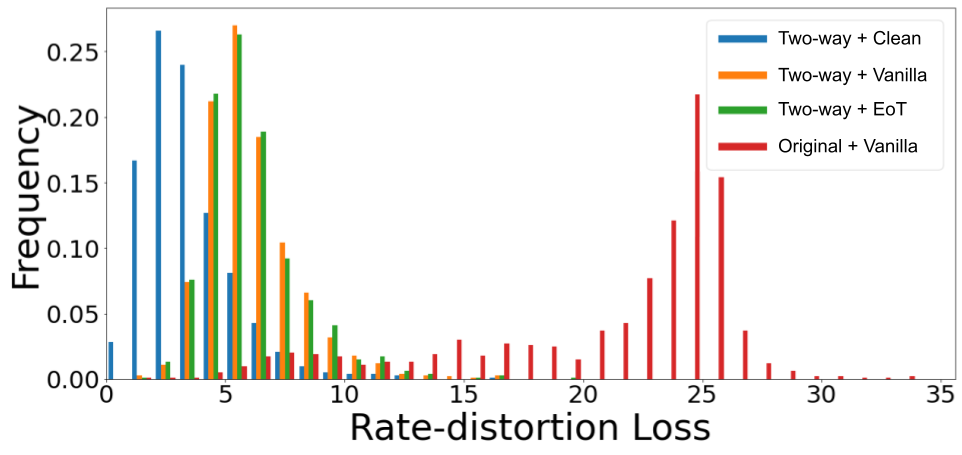}
    \vspace{-2mm}
    \caption{
        Rate-distortion loss histogram for test samples under rate attacks.
    }
    \vspace{-2mm}
    \label{fig:histogram}
\end{figure*}

Figure~\ref{fig:histogram} visualizes the rate-distortion loss value distribution of test samples for the highest bitrate models tested in the experiments for Figure~\ref{fig:defense_main1}(a) of the main paper.
Note that the results of our method exhibit low losses by avoiding extreme failure cases with high probability.

\section{Comparison to Hand-crafted Codecs}
\begin{figure*}[t]
    \centering
    \includegraphics[width=0.8\linewidth]{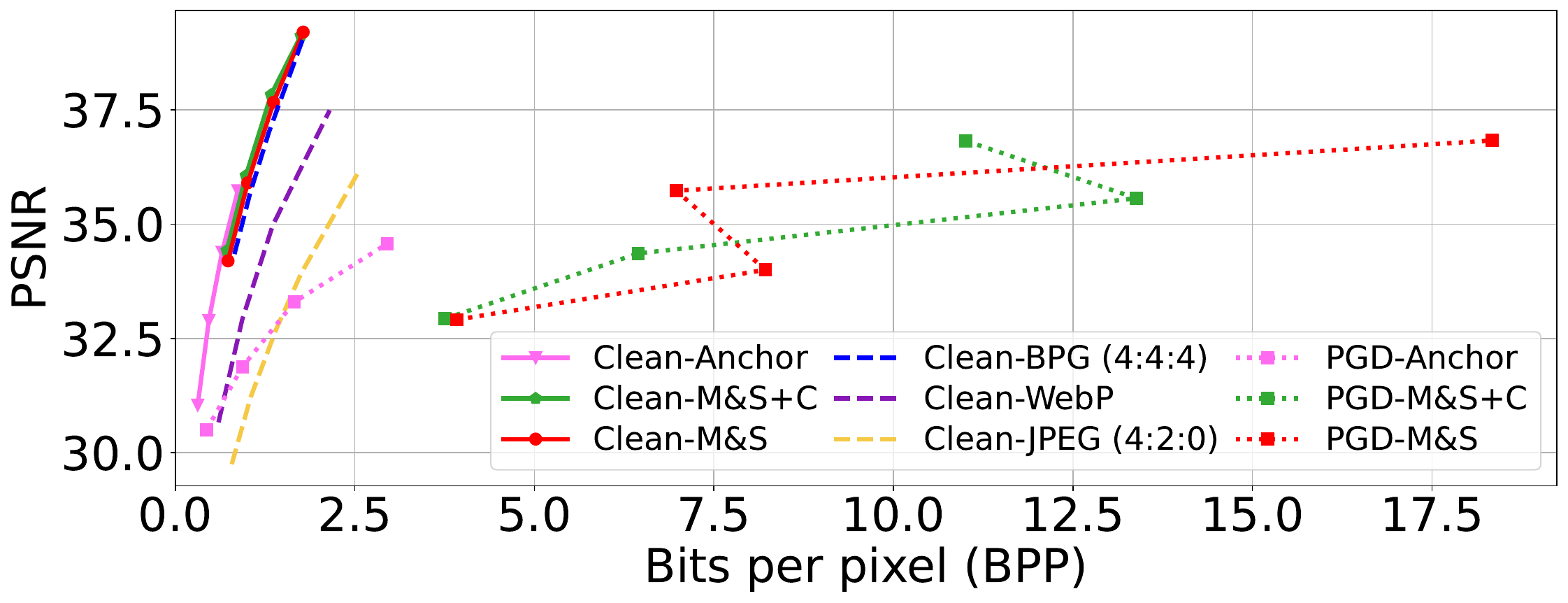}
    \vspace{-2mm}
    \caption{
        Rate-distortion results of attacked learned image compression models and traditional codecs.
    }
    \vspace{-2mm}
    \label{fig:vs_handcrafted}
    \vspace{3mm}
\end{figure*}

Figure~\ref{fig:vs_handcrafted} compares the compression performance between the attacked models and hand-crafted codecs.
We observe the severe performance degradation of the attacked models, which is even worse than the hand-crafted codecs.

